\documentclass[aps,prl,preprint,showpacs,groupedaddress]{revtex4}

\begin{document}


\title{Orbital Disordering and metal-insulator transition\\ 
with hole-doping in perovskite-type vanadium oxides}

\author{J. Fujioka$^{1}$, S. Miyasaka$^{1}$, and Y. Tokura$^{1,2,3}$}
\affiliation{$^{1}$Department of Applied Physics, University of Tokyo, Tokyo 113-8656, Japan}
\affiliation{$^{2}$Spin Superstructure Project, ERATO, Japan Science and Technology Agency, 
Tsukuba, 305-8562, Japan}
\affiliation{$^{3}$Correlated Electron Research Center (CERC), National Institute of Advanced 
Industrial Science and Technology (AIST), Tsukuba 305-8562, Japan}

\date{\today}

\begin{abstract}

Filling-control metal-insulator transitions (MITs) and 
related electronic phase diagrams have been 
investigated for hole-doped vanadium oxides, 
Pr$_{1-x}$Ca$_x$VO$_3$, Nd$_{1-x}$Sr$_x$VO$_3$ and Y$_{1-x}$Ca$_x$VO$_3$, 
with perovskite structure. 
The increase of the doping level $x$ causes the melting of the $G$-type (and 
$C$-type) orbital order, prior to or concomitantly with the MIT, due partly
to the doped-hole motion and partly to the ramdom potential arising from
the quenched disorder. 
In particular, the $G$-type spin- and $C$-type orbital-ordered phase present
in Y$_{1-x}$Ca$_x$VO$_3$ disappears immediately upon hole doping, 
around $x=0.02$. On the other hand, 
the critical doping level $x$ for MIT is governed by the electron-correlation 
strength of the undoped parent compound.

\end{abstract}



\maketitle


The orbital degree of freedom in $3d$ transition-metal compounds has been 
attracting much attention, since it plays an important role not only in 
producing versatile magnetic structures but also in dramatically modefying 
charge transport as observed in the metal-insulator transition (MIT) and
the colossal magnetoresistance phenomena\cite{RevModPhys,science}. 
In particular, recent experimental and theoretical studies on perovskite-type 
manganites have clarified the close interplay among orbital, spin, charge 
and lattice degrees of freedom\cite{CMR}. 
In manganese oxides the strong Jahn-Teller effect of 
the partially filled $e_g$-orbital 
dominates the lattice-coupled charge dynamics, as manifested by, 
for example, the charge/orbital ordering and disordering accompanying the 
metal-insulator phenomena. 
In the $t_{2g}$-orbital sector, by contrast, the orbital-lattice interaction 
is much weaker 
than that in the $e_g$ ones. 
Therefore, the interplay between orbitals and spins, 
not only the intersite exchange interaction but also the intra-atomic 
spin-orbit interaction, is more visible in vanadium oxides with active 
$t_{2g}$-orbital electrons\cite{RVO3,LaVO3}.

A recent investigation on La$_{1-x}$Sr$_x$VO$_3$ has revealed the critical 
role of the $t_{2g}$ orbital and spin correlations 
near the boundary of the MIT\cite{LSVO}. 
The parent compound LaVO$_3$ with perovskite structure 
is a prototypical Mott-Hubbard insulator, 
where V is nominally trivalent and has 3$d^2$ 
configuration, that is, two valence electrons in the 3$d$ orbitals ($t_{2g}$ 
manifold) with spin S=1. 
Since the orthorhombic distortion lifts the degeneracy of the energy level 
of $t_{2g}$ orbitals, one electron always 
occupies the $d_{xy}$ orbital and the other one either $d_{yz}$ 
or $d_{zx}$ orbital. With lowering temperature ($T$), LaVO$_3$ 
undergoes two successive phase transitions\cite{Zubkov, Bordet}. 
First, the magnetic transition from paramagnetic (PM) to 
$C$-type spin ordered state, where spins align ferromagnetically 
along the $c$ axis and stagger in the $ab$ plane, occurs at 
$T_{\mathrm{SO1}}$=143 K. 
Subsequently, the structural phase transition accompanying the 
$G$-type orbital ordering (OO), 
where $d_{yz}$ and/or $d_{zx}$ orbitals stagger 
in all ($x,y,z$) directions, occurs at $T_{\mathrm{OO1}}$=141 K 
as shown Fig.1(a) \cite{RVO3}. 
By partially replacing La with Sr, which results in hole doping 
(or decreasing band-filling) , 
the filling-control MIT can be achieved. 
The sequential order and causality of the $C$-type spin ordering 
(SO) and the $G$-type OO 
are also observed in La$_{1-x}$Sr$_x$VO$_3$ in an insulating 
region ($x$$<$0.178). The MIT occurs around $x$=0.178 for 
La$_{1-x}$Sr$_x$VO$_3$ accompanying the melting of the $G$-type OO 
and related structural phase transition, while the $C$-type SO remains 
up to around $x$=0.260 and forms the antiferromagnetic metallic state.

When La is replaced with other rare-earth elements ($R$) and accordingly the 
orthorhombic-lattice distortion of the perovskite structure is changed, 
many of $R$VO$_3$ show different $T$-dependent 
sequential order of the SO and OO transitions\cite{RVO3}. 
Additionally, some of them bear the different pattern of SO and OO, 
i.e. the $G$-type SO and $C$-type OO as the ground state. 
The whole spin-orbital phase diagram 
of $R$VO$_3$\cite{RVO3} is reproduced in Fig.1(a). In $R$VO$_3$ with 
$R$=Pr to Lu, the $G$-type OO appears well above $T_{\mathrm{SO1}}$ and 
accordingly the $G$-type orbital-ordered but PM phase exists 
in the phase diagram. Thus, the sequential order and causality 
concerning the $C$-type SO and the $G$-type OO 
are opposite between $R$VO$_3$ with $R$=Pr to Lu and those 
with $R$=La and Ce. 
As the typical examples of $R$VO$_3$ showing the OO 
first, PrVO$_3$ and NdVO$_3$ are to be investigated here. 
In $R$VO$_3$ with $R$=Dy to Lu, 
on the other hand, the $G$-type SO and $C$-type OO 
appear in the low-$T$ region (see Fig. 1(a)). 
Among them, the spin- and orbital-ordered state for YVO$_3$ has been studied 
as the prototype experimentally\cite{Blake, Noguchi, Ulrich, Kawano} 
and theoretically\cite{Khaliullin, Sawada}. 
As in La$_{1-x}$Sr$_x$VO$_3$, the hole doping by partially replacing 
the trivalent $R$ with the divalent alkaline earth ones ($A$=Sr or Ca), 
causes the MIT. We have chosen Pr$_{1-x}$Ca$_x$VO$_3$, 
Nd$_{1-x}$Sr$_x$VO$_3$ and Y$_{1-x}$Ca$_x$VO$_3$ as the hole 
doped systems for PrVO$_3$, NdVO$_3$ and YVO$_3$, respectively. 
To clarify the behavior of SO and OO in the hole doped region 
and the critical behavior of OO in the vicinity of MIT for 
these compounds, we have prepared the single crystals by a floating 
zone method\cite{LSVO} and systematically investigated 
transport, specific heat and magnetization with varying $x$.

The $T$-dependence of the resistivity $\rho$ for 
Pr$_{1-x}$Ca$_x$VO$_3$ is shown in Fig. 2. 
The $\rho$ for 0$\le$$x$$\le$ 0.23 shows an insulating behavior 
and the extrapolated zero-$T$ conductivity remains zero, 
while finite for $x$$>$0.25. This indicates that the filling-control MIT 
at zero-$T$ seems to occur at $x\sim0.25$. 
The critical doping level $x_c$ for the MIT is similarly 
determined for Nd$_{1-x}$Sr$_x$VO$_3$ ($x_c$$\sim$$0.23$). 
The $x_c$(=0.5) for Y$_{1-x}$Ca$_x$VO$_3$ was reported in 
ref.\cite{Kasuya}. In general, the partial substitution of 
$R$ with $A$ changes not only the nominal hole concentration but also 
the change of crystal structure 
and accordingly leads to the change of the effective one-electron bandwidth. 
To see how the change of the effective one-electron bandwidth 
affects the MIT, we plotted the tolerance factor of each compound 
as a function of the doping level as shown in Fig.1 (b). 
In this regime, the tolerance factor represents the relative 
one-electron bandwidth\cite{okimoto}. 
As seen in Fig. 1(b), $x_c$ increases systematically 
with decreasing the tolerance factor, or equivalently with decreasing 
the effective one-electron bandwidth as observed for
$R_{1-x}A_x$TiO$_3$\cite{Katsufuji1}.
This is also consistent with the theoretical results\cite{RevModPhys,Imada}.

The $T$-dependence of the specific heat and magnetization 
for Y$_{1-x}$Ca$_x$VO$_3$, Nd$_{1-x}$Sr$_x$VO$_3$, and 
Pr$_{1-x}$Ca$_x$VO$_3$ is shown in Fig. 3. 
In the respective end ($x$=0) compounds, 
specific heat curves show two or three peaks 
corresponding to magnetic transitions and/or structural ones coupled 
with OO, as previously reported\cite{RVO3}. 
The peaks at $T_{\mathrm{OO1}}$ and $T_{\mathrm{SO1}}$ correspond to the onset 
of the $G$-type OO and the $C$-type SO, respectively. 
The specific-heat jump observed for YVO$_3$ at 
$T_{\mathrm{SO2}}$ (=$T_{\mathrm{OO2}}$) 
corresponds to the first-order transition to 
the $G$-type spin- and the $C$-type orbital-ordered phase. 
The magnetization curve also shows upturn at $T_{\mathrm{SO1}}$ and jump 
at $T_{\mathrm{SO2}}$, while no anomaly is observed at $T_{\mathrm{OO1}}$. 
Also in the Ca- or Sr- doped compounds the specific heat 
and magnetization curves show anomalies corresponding to 
the appearance of SO and OO. 
At first, we focus on Y$_{1-x}$Ca$_x$VO$_3$ among these three compounds. 
For $x$=0.01, the specific heat curve shows three discontinuities as seen 
in the end compound ($x$=0). For $x$=0.02, however, the anomalies 
corresponding to the transition to the $G$-type spin- and $C$-type 
orbital-ordered phase cannot be observed down to 2 K, 
while the other two peaks can be clearly as well. 
This indicates that the $G$-type spin- and $C$-type orbital-ordered 
phase is fragile against the hole doping and disappears at such a low 
doping level as $x$=0.02. 
With increasing $x$, the peak corresponding to 
the onset of the $G$-type OO gradually broadens, shifts to lower 
$T$, and above $x$=0.10 is hardly discerned. Thus, the correlation 
of the $G$-type OO becomes weaker with increasing $x$ and the $G$-type OO 
finally melts around $x$=0.10. With increasing $x$, 
the peak corresponding to the transition to the $C$-type SO also 
gradually broadens and consequently no peak can be discerned for $x$=0.11. 
The magnetization curve, however, shows upturn around 100 K, 
indicating that the $C$-type SO still exists in this compound 
as shown in Fig. 3(d). This upturn can be seen up to $x$$\sim$0.6, 
and consequently the $C$-type SO seems to exist even 
in a metallic region as in the case of La$_{1-x}$Sr$_x$VO$_3$.
These doping-dependent features are summarized in Fig. 4 as the electronic
phase diagram.

As shown in Figs. 3(b), (c), (e) and (f), the specific heat curves 
for Nd$_{1-x}$Sr$_x$VO$_3$ and Pr$_{1-x}$Ca$_x$VO$_3$ show 
similar behaviors to those for Y$_{1-x}$Ca$_x$VO$_3$ apart from 
the difference in the ground state at $x$=0. For Nd$_{1-x}$Sr$_x$VO$_3$ and 
Pr$_{1-x}$Ca$_x$VO$_3$, the peak corresponding to 
the onset of the $G$-type OO at $T_{\mathrm{OO1}}$ disappears around $x$=0.10 
and $x$=0.20, respectively, indicating the melting of the $G$-type OO. 
It is noted that the peak for Pr$_{1-x}$Ca$_x$VO$_3$ can be observed 
up to a larger $x$ value than those for Y$_{1-x}$Ca$_x$VO$_3$ 
and Nd$_{1-x}$Sr$_x$VO$_3$, indicating that the long-range $G$-type OO 
remains up to a higher doped region for Pr$_{1-x}$Ca$_x$VO$_3$. 
The magnetization curve also shows upturn corresponding to the onset of 
the $C$-type SO at $T_{\mathrm{SO1}}$ up to $x$=0.26 for 
Nd$_{1-x}$Sr$_x$VO$_3$ and up to $x$=0.30 for Pr$_{1-x}$Ca$_x$VO$_3$ 
(see Fig.4), respectively.

The $T$-dependence of $\rho$ in a small $x$ region 
for Pr$_{1-x}$Ca$_x$VO$_3$ also shows broad kinks 
which perhaps reflect the transition to the $G$-type OO and the $C$-type SO 
as shown in Fig. 2. This contrasts with the clear kink 
due to the first-order phase transition observed in La$_{1-x}$Sr$_x$VO$_3$
\cite{LSVO}. 
To see the anomaly more clearly, the $T$-derivative of 
logarithmic $\rho$ ($d$log$\rho$/$dT$) is shown in the inset of Fig. 2. 
For $x$=0.10, it shows two dips around 
$T_{\mathrm{OO1}}$ and $T_{\mathrm{SO1}}$. 
For $x$=0.25, only the broad dip corresponding to the onset of the $C$-type SO 
can be seen around $T_{\mathrm{SO1}}$. The evolution of the $C$-type SO 
seems to supress the charge transport effectively, 
since $d$log$\rho$/$dT$ increases steeply 
below $T_{\mathrm{SO1}}$ for the both compounds .

The electronic phase diagram, obtained by plotting the transition 
temperatures of the SO and OO as a function of $x$, is shown for each compound 
in Fig. 4, which includes that of La$_{1-x}$Sr$_x$VO$_3$ 
previously reported\cite{LSVO}. 
For Pr$_{1-x}$Ca$_x$VO$_3$, Nd$_{1-x}$Sr$_x$VO$_3$ 
and Y$_{1-x}$Ca$_x$VO$_3$, $T_{\mathrm{OO1}}$ and $T_{\mathrm{SO1}}$ 
systematically decrease with increasing $x$ and the $T$ 
interval between them also decreases for these three compounds. 
As mentioned above, the $G$-type OO seems to melt or at least 
become obscure around $x$=0.10 
for Y$_{1-x}$Ca$_x$VO$_3$, $x=0.09$ for Nd$_{1-x}$Sr$_x$VO$_3$ and 
$x$=0.20 for Pr$_{1-x}$Ca$_x$VO$_3$, 
respectively. Thus, the doping level where the $G$-type OO melts 
(defined as $x_{\mathrm{o}}$ here) is smaller than $x_c$ for the MIT. 
This also contrasts with that 
the $G$-type OO remains until the MIT, i.e. $x_o$=$x_c$, 
in La$_{1-x}$Sr$_x$VO$_3$ with $T_{\mathrm{SO1}}$ locating always 
above $T_{\mathrm{OO1}}$. 
If the hole motion alone could cause the collapse of the $G$-type OO, 
$x_{\mathrm{o}}$ would decrease 
with the increase of the effective one-electron bandwidth, 
i.e. the kinetic energy of the doped holes. 
However, $x_{\mathrm{o}}$ does not 
change systematically with the tolerance factor, or equivalently
the effective one-electron bandwidth, in contrast to the case of $x_c$ 
for the MIT. 
Thus, there seems to be another factor which destabilizes 
the PM and $G$-type orbital-ordered phase in addition to 
the increase of the kinetic energy of holes. The most plausible one is 
the quenched disorder arising from the solid solution 
of $R$ (small) and $A$ (large) ions, 
as recently demonstrated by the investigation 
on manganese oxides\cite{Attfield, Tomioka}. 
To see the effect of the quenched disorder 
it is reasonable to compare Nd$_{1-x}$Sr$_x$VO$_3$ with 
Pr$_{1-x}$Ca$_x$VO$_3$. These systems 
share the close value of the one-electron bandwidth 
in the doping level of $x$$\sim$0.2, where the melting of the $G$-type OO 
is observed (see Fig.1(b)). We defined the variance of the $R$/$A$ ionic radii 
as a function of $x$, $\sigma^2$=$\Sigma (x_ir_i^2-r_A^2)$, 
as the measure of the magnitude of the quenched disorder\cite{Attfield}. 
Here, $x_i$, $r_i$ and $r_A$ are the fractional occupancies, the effective 
ionic radii of cations of $R$ and $A$, and the averaged ionic radius 
($r_A$=$(1-x)r_{\mathrm{R}^{3+}}+xr_{\mathrm{A}^{2+}}$), respectively. 
At $x$=0.10, $\sigma^2$ for Nd$_{1-x}$Sr$_x$VO$_3$ is about 
2.6$\times$$10^{-5}$ nm$^2$, 
whereas that for Pr$_{1-x}$Ca$_x$VO$_3$ 3.2$\times$$10^{-6} $nm$^2$ 
is one order of magnitude smaller. 
Since the effective one-electron bandwidth for Nd$_{1-x}$Sr$_x$VO$_3$ 
is close to that for Pr$_{1-x}$Ca$_x$VO$_3$ in this doping level, 
this suggests that as well as the motion of the doped hole the increase of 
the magnitude of the quenched disorder results in the disappearance of 
the long-range $G$-type OO. This is perhaps because the disorder randomizes 
the local lattice distortion 
which would be induced by the collective Jahn-Teller coupling 
distortion concomitant with the $G$-type OO, and hence plays a role of random 
field acting on the orbital (pseudo-spin) sector. 
To see the lattice distortion coupled with the $G$-type OO, 
Raman-scattering spectroscopy can be used because of its sensitivity 
to the lattice distortion. For NdVO$_3$, an activated Raman mode is observed 
at around 700cm$^{-1}$ below $T_{\mathrm{OO1}}$\cite{RVO3}. 
With increasing $x$ the peak broadens and its integrated intensity 
also decreases. This is indicative of the suppression of 
the lattice distortion coupled with the $G$-type OO. At $x=0.12$ 
for Nd$_{1-x}$Sr$_x$VO$_3$ where the long-range 
PM and $G$-type orbital-ordered phase 
seems to be absent, a broad peak-structure can be observed 
in a low-$T$ region, suggesting the subsistence of the short range 
correlation of the $G$-type OO. Quite a similar disorder-induced melting 
of the long range OO into the short-range correlation has also 
been reported for manganites\cite{Tomioka2}. 
On the other hand, the $C$-type SO 
is robust against the increase of $x$ and remains 
even in the metallic region, which is similar to 
the case of La$_{1-x}$Sr$_x$VO$_3$. The effect of the quenched disorder 
is much less on the $C$-type SO than on the $G$-type OO, perhaps reflecting 
the smaller coupling of SO with the lattice. 
Finally, some remarks should be added about the doping effect on 
the $G$-type spin- and the $C$-type orbital-ordered phase 
in Y$_{1-x}$Ca$_x$VO$_3$. As observed in Fig. 4, the $G$-type spin- and 
$C$-type orbital-ordered phase is extremely unstable against 
such a small hole doping 
as $x$=0.02. The magnitude of the quenched disorder is supposed to be 
minimal in such a lightly doped region. 
A recent theoretical calculation\cite{Ishihara2} 
predicts that the motion of the doped hole reduces 
the spin order parameter, which causes the softening of orbiton and 
leads to the instability of the $G$-type spin- and 
the $C$-type orbital-ordered phase. The extreme sensitivity to the doping 
level may also arise from the bicritical phase competition with the adjacent 
$G$-type OO and $C$-type SO phase that governs the ground state phase for 
0.02$<$$x$$<$0.10.

In summary, we have revealed the electronic phase diagram 
for Pr$_{1-x}$Ca$_x$VO$_3$, Nd$_{1-x}$Sr$_x$VO$_3$ and Y$_{1-x}$Ca$_x$VO$_3$ 
by the transport, specific heat 
and magnetization measurements. The paramagnetic and long-range $G$-type 
orbital ordered phase is unstable against the increase of 
the doping level $x$, which is attributed not only to 
the hole motion but also to the increase 
of the quenched disorder arising from the random chemical substitution of 
the rare-earth elements with alkaline earth ones. 
On the other hand, 
the critical doping level for the insulator-metal transition 
is governed by the one-electron bandwidth. Moreover, the $G$-type spin- and 
the $C$-type orbital-ordered phase for Y$_{1-x}$Ca$_x$VO$_3$ disappears 
at a minimal doping as low as $x$=0.02 and, 
hence, is extremely sensitive to the change of band-filling, as 
compared with the $C$-type spin- and $G$-type 
orbital-ordered one.

We would thank S. Ishihara for helpful discussion.



\newpage

 \begin{figure}[htbp!]
 \caption{
(a) The spin-orbital phase diagram in $R$VO$_3$($R$=Lu-La). 
Closed and open circles, and open triangle indicate the 
transition temperatures of the $G$-type orbital ordering (OO), 
$T_{\mathrm{OO1}}$,  
the $C$-type spin ordering (SO), $T_{\mathrm{SO1}}$, 
and the $G$-type SO and the $C$-type OO, 
$T_{\mathrm{SO2}}=T_{\mathrm{OO2}}$, respectively. 
Schematic representations of 
the $G$-type OO, the $G$-type SO and $C$-type OO, 
and the $C$-type OO and $G$-type OO are also shown, in which 
open arrows and lobes indicate spins, and occupied $d_{yz}$ and $d_{zx}$
orbitals on the vanadium ions, respectively. 
(b) The doping level versus the tolerance factor 
for La$_{1-x}$Sr$_x$VO$_3$, Pr$_{1-x}$Ca$_x$VO$_3$, Nd$_{1-x}$Sr$_x$VO$_3$ 
and Y$_{1-x}$Ca$_x$VO$_3$. Open circles indicates the critical 
doping level $x_c$ of the metal-insulator transition.}
 \label{fig1}
 \end{figure}

 \begin{figure}[htbp!]
 \caption{Temperature dependence of resistivity for single crystals 
of Pr$_{1-x}$Ca$_x$VO$_3$ with various $x$. 
The inset shows the temperature derivative of 
logarithmic resistivity ($d$log$\rho$/$d$T). 
The closed and open triangles indicate the transition temperatures of 
the $G$-type OO and the $C$-type SO, respectively.}
 \label{fig2}
 \end{figure}

\begin{figure}[htbp!]
 \caption{The temperature dependence of specific heat and magnetization for 
Y$_{1-x}$Ca$_x$VO$_3$ ((a),(d)), Nd$_{1-x}$Sr$_x$VO$_3$ ((b),(e)), 
and Pr$_{1-x}$Ca$_x$VO$_3$ ((c),(f)). 
The closed, open and double triangles indicate the transition 
temperatures of the $G$-type OO, the $C$-type SO, and 
the $G$-type SO and $C$-type OO, respectively.}
 \label{fig3}
 \end{figure}

 \begin{figure}[htbp!]
 \caption{The electronic phase diagram as a function of the doping level $x$
for La$_{1-x}$Sr$_x$VO$_3$\cite{LSVO}, 
Pr$_{1-x}$Ca$_x$VO$_3$, Nd$_{1-x}$Sr$_x$VO$_3$ 
and Y$_{1-x}$Ca$_x$VO$_3$. The open, closed circles, 
and open triangles indicate the transition temperatures of 
the $G$-type OO ($T_{\mathrm{OO1}}$), the $C$-type SO ($T_{\mathrm{SO1}}$) 
and the $G$-type SO and the $C$-typeOO ($T_{\mathrm{OO2}}=T_{\mathrm{SO2}}$), respectively.}
 \label{fig4}
 \end{figure}

\end{document}